\documentclass[11pt]{winnower}

\usepackage{wasysym} 
\usepackage{amsmath,color} 
\usepackage[normalem]{ulem} 

\begin{document}

\newcommand{\re}[1]{{#1}}

\newcommand{\solrad}{\ifmmode{R}_{\rm S}\else${R}_{\rm S}$\fi}
\newcommand{\solmas}{\ifmmode{M}_{\rm S}\else${M}_{\rm S}$\fi}
\newcommand{\ergu}{$\,$ergs$\,$s$^{-1}$}
\newcommand{\tintu}{\ifmmode{\rm erg~cm^{-2}~s^{-1}sr^{-1}}\else
  erg~cm$^{-2}$~s$^{-1}$~sr$^{-1}$\fi}

\newcommand{\intu}{\ifmmode{\rm erg~cm^{-2}~s^{-1}sr^{-1}Hz^{-1}}\else
  erg~cm$^{-2}$~s$^{-1}$~sr$^{-1}$Hz$^{-1}$\fi}

\newcommand{\fluxu}{\ifmmode{\rm erg~cm^{-2}~s^{-1}}\else
  erg~cm$^{-2}$~s$^{-1}$\fi}

\newcommand{\fluxau}{\ifmmode{\rm erg~cm^{-2}~s^{-1}Hz^{-1}}\else
  erg~cm$^{-2}$~s$^{-1}$Hz$^{-1}$\fi}

\newcommand{\photau}{\ifmmode{\rm ph~cm^{-2}~s^{-1}Hz^{-1}}\else
  ph~cm$^{-2}$~s$^{-1}$Hz$^{-1}$\fi}

\newcommand{\velu}{$\,$km$\,$s$^{-1}$}
\newcommand{\dynu}{$\,$dyn$\,$cm$^{-2}$}
\newcommand{\wave}{\ifmmode{\lambda} \else$\lambda$\fi}
\newcommand{\freq}{\omega}

\newcommand{\avg}[1]
%{\ifmmode\overline{#1}
%\else   $\overline{#1}$\fi}
{\ifmmode \langle {#1} \rangle_\tau \else $\langle {#1} \rangle_\tau$
  \fi}

\newcommand\figboth{
\begin{figure} 

\includegraphics[width=0.52\linewidth]{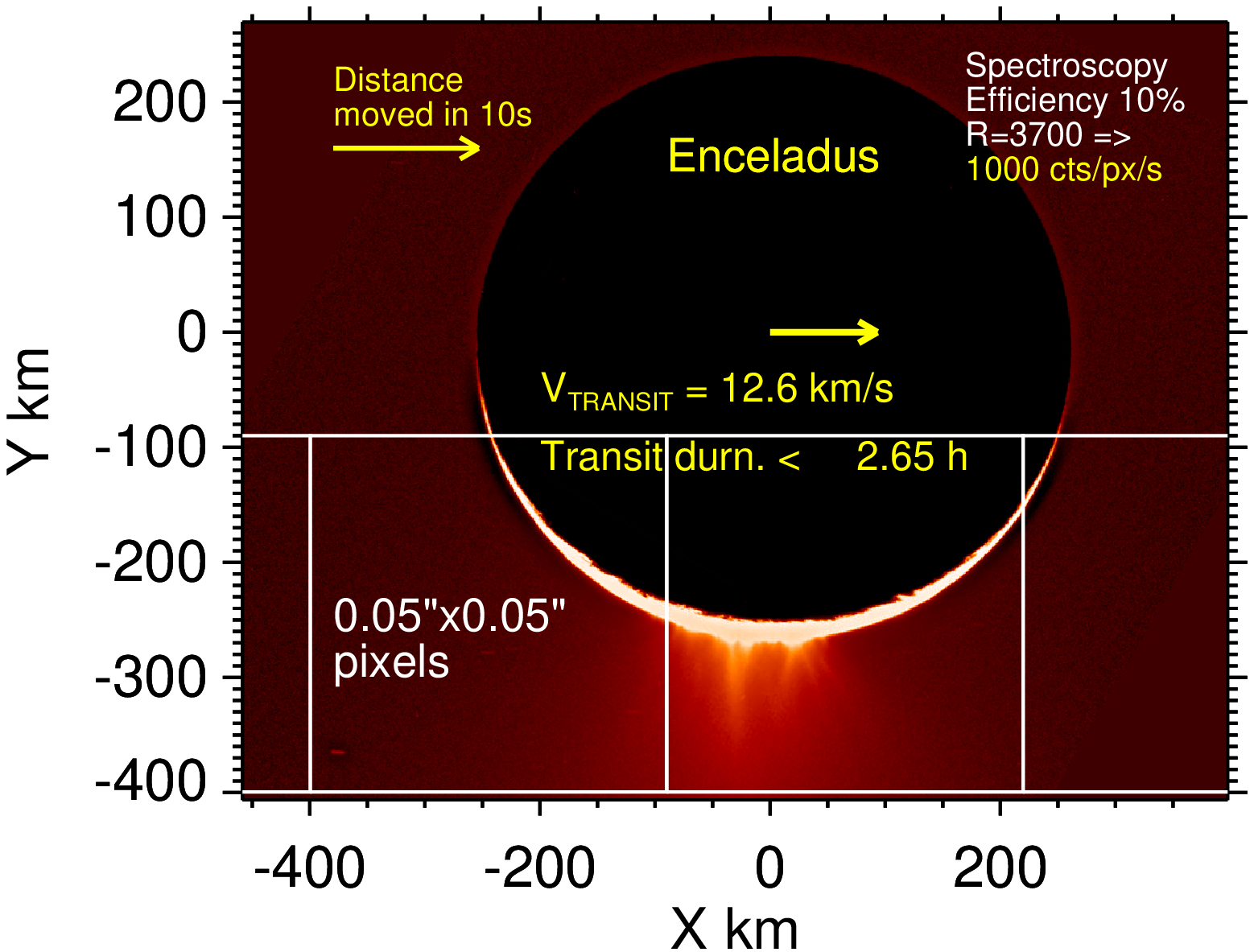}
\includegraphics[width=0.52\linewidth]{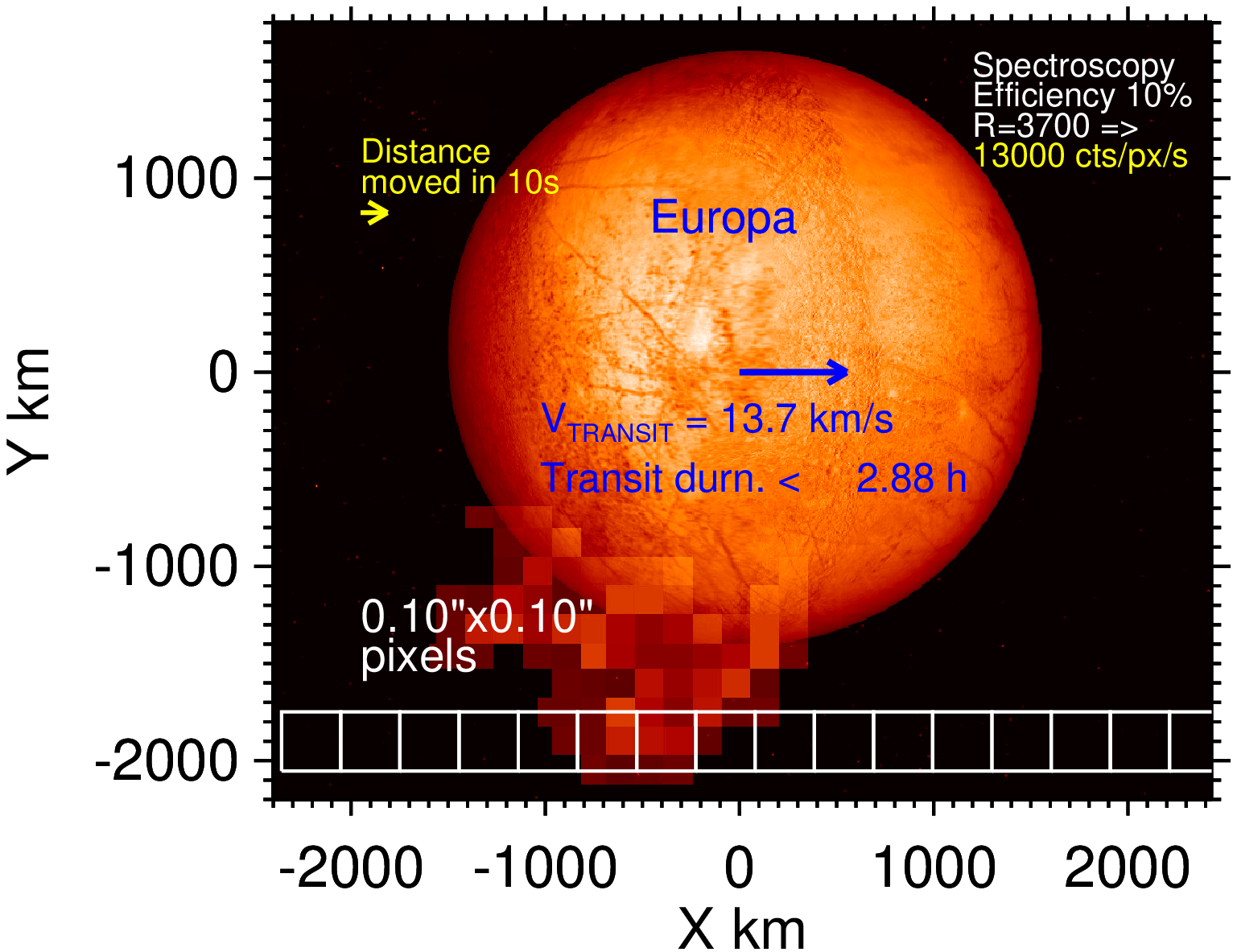}
\caption{
\label{fig:both}
Images of Enceladus (left) and Europa (right) are superposed with data
relevant to transits. The images are from NASA's websites.    }
\end{figure}
}

\newcommand\figspec{

\begin{figure}
\includegraphics[width=1.0\linewidth]{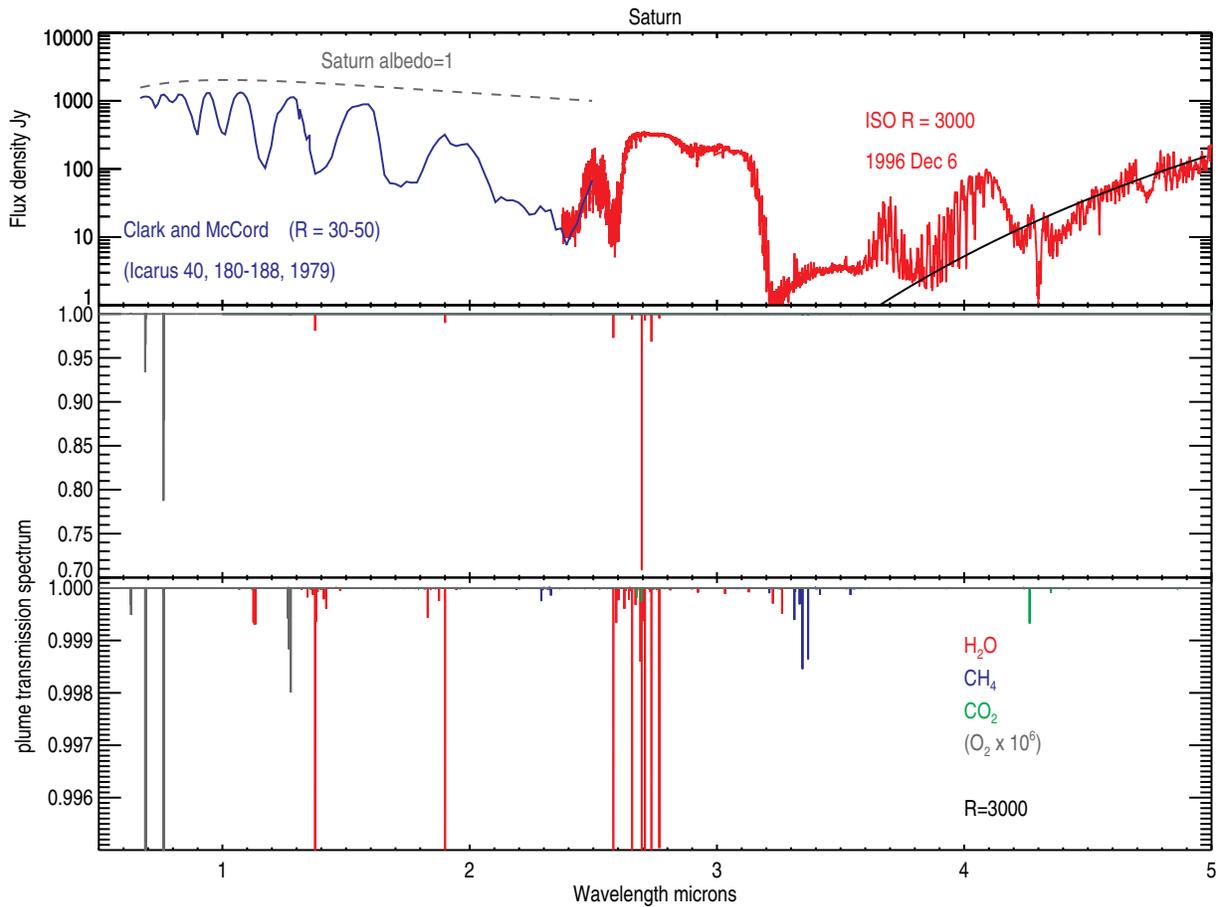}
\caption{\label{fig:spec} A composite visible-infrared spectrum of
  Saturn is shown in the top panel. The dashed line shows the
  brightness of Saturn with a uniform albedo of 1, and the blue line
  shows a low spectral resolution IR spectrum from
  \cite{Clark+McCord1979}.  The red curve shows a spectrum from the
  ISO satellite obtained from the ISO data archive which has a
  spectral resolution of about 3000. The solid black line is a black
  body flux spectrum at 174 K from Saturn. The lower panels show
  spectral transmission calculations for the plumes of Enceladus, scaled differently to reveal strong and weak transitions,
  computed using the HITRAN database \cite{Rothman+others2013} using a spectral resolution of
  3000. Other details of the calculations are given in the text. }
\end{figure}
}

\newcommand\figpos{

\begin{figure*}
\includegraphics[width=1.0\linewidth]{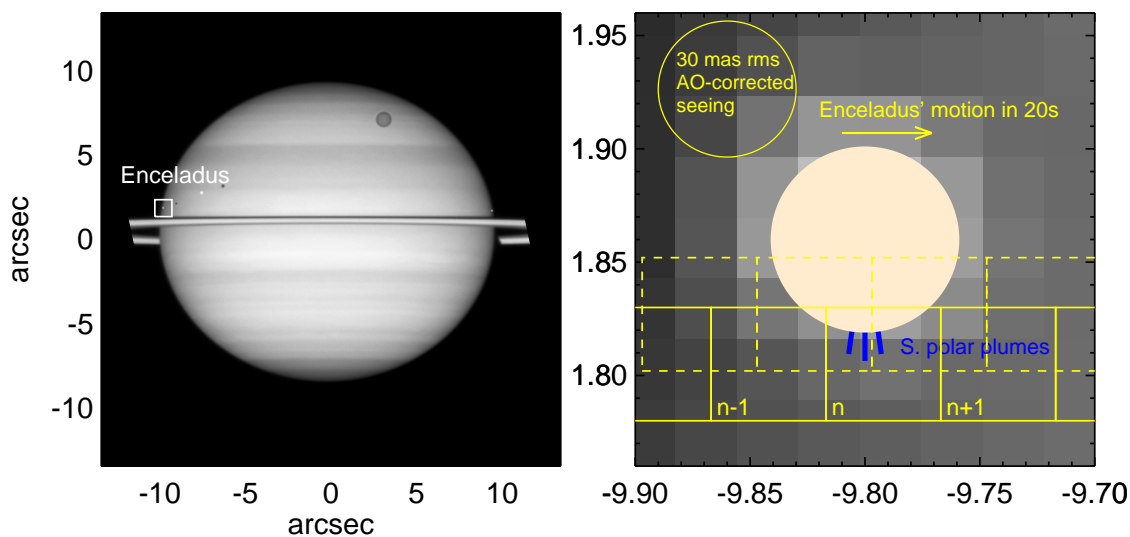}
\caption{\label{fig:pos} An image of Saturn obtained by the {\em
    Hubble Space Telescope} is shown on the left, acquired on February
  24th 2009 and obtained from NASA's APOD website {\small \tt
    https://apod.nasa.gov/apod/ap090319.html}.  Four of Saturn's
  satellites are captured in various stages of transiting the planet's
  disk.  On the right is a blow up from the small region within the
  square area in the left panel containing Enceladus (shown as a sharp
  opaque almond-colored disk superimposed on the highly expanded
  Hubble image underneath).  On this panel (spanning merely 0.2
  seconds of arc on a side) a 30 milli-arc-second radius circle is
  shown representing \re{(optimistically) a good 8m telescope point spread function operating 
at $\lambda= 1$ micron}, after
  adaptive optical correction.  The boxes are $0.05\times0.05"$ areas at a
  hypothetical spectrograph slit oriented E-W on the sky.  The dashed
  box shows a $1\sigma$ residual image motion of 30 milli-arcseconds
  on the sky. (In reality the images will move relative to the fixed
  spectrograph slit, it is shown in this manner for clarity.)  }
\end{figure*}
}

\newcommand\tabone{
\begin{table}[h]
\begin{center}
 \caption{Circumstances of transits and plumes at Enceladus and
   Europa, at opposition
\label{tab:circ}    
}
 \begin{tabular}{llrr} 
 \hline \hline Parameter & Unit & Enceladus & Europa\\ \hline \\ Mean
 distance (opposition) & km & 1.278(9) & 6.39(8) \\ radius &km & 257 &
 1480\\ apparent diameter &arcsec & 0.081" & $0.97"$\\ 
cross-plume column density
  & particles cm$^{-2}$ & $1.5\times10^{16}$  & $\ldots$ \\ plume scale
 length & km & $\approx 100$ & $\approx 500$ \\ 
plume scale length
 &arcsec & $\approx 0.016"$ & $\approx 0.16"$ \\ mean orbital speed
 &km/s & 12.63 & 13.74\\ radius/orbital speed &seconds & 20.35 &
 107.7\\ plume scale length/orbital speed &seconds & 8 & 36\\ maximum
 transit duration &hours & 2.65 & 2.88\\ Earliest next transit date &&
 Spring of 2022 & \ldots \\ \\ \hline
\end{tabular}
\end{center}
{Data are standard sources, some computed using the JPL ephemeris,  and, for plumes, references in the text. }
\end{table}
}

\newcommand\tabobs{
\begin{table}
\begin{center}
 \caption{A comparison of three large infrared observatories
\label{tab:obs}    }
 \begin{tabular}{lllll} 
 \hline \hline Parameter & Unit & DKIST & JWST & Keck II\\ \hline
 \\ Primary aperture &m & 4 & 6.5 & 10 \\ Operations & & 2019- & 2018-
 & 1996- \\ Diffraction limit at $1\mu$m & mas$^\ast$ & 63 & 39 &
 25\\ IR spectrograph & & CRYO-NIRSP$^+$  & NIRSpec & NIRC-2
 Grisms\\ 
Minimum pixel size & mas & 150& 100 & 10-40\\
${\cal R}= \lambda/\Delta\lambda$ & & 30,000 & 2,700 &
 2,500-11,000\\ AO Strehl ratio$^\dag$ & & 0.3-0.6$^\$$ & \ldots &
 0.35$^\ddag$\\ Image stability & mas & \ldots & $< 3.7$&
 \ldots\\ 
maximum slew rate & mas/sec & \ldots & $\leq 30$ & \ldots \\
Other & & Coronagraph, & L2 orbit, obervations & \\ 
 & & polarimetry & limited to near quadrature$^\#$ & \\ 
\hline
\end{tabular}
\end{center}
{$^\ast$Milli-arc-seconds.  $^\dag$The Strehl ratio is defined as the
  peak intensity of a point source divided by the peak intensity of
  the (theoretical) diffraction-limited
  point spread function (PSF). If the PSF's have a similar shape, then the rms seeing disk is of order the inverse of the Strehl ratio larger than diffraction.  $^+$\cite{Fehlmann+others2016}. $^\$$\cite{Johnson+others2014}.
  $^\ddag$\cite{VanDam+others2006}. $^\#$\cite{Keszthelyi+others2016}.
NIRSpec and NIRC-2 data are from instrument web pages, 
{\small https://jwst.nasa.gov/nirspec.html and  https://www2.keck.hawaii.edu/inst/nirc2/genspecs.html}. Note that the 30 mas/sec maximum slew rate for JWST at Jupiter corresponds to $\approx 110$ km~s$^{-1}$ at the planet.
}
\end{table}
}

\title{Hypothesis article: A novel strategy to seek bio-signatures at Enceladus and Europa}

\author{Philip Judge}

\affil{HAO, NCAR\footnote{The National Center for Atmospheric Research is supported by the National Science Foundation}, Box 3000, Boulder CO 80307, USA.  judge@ucar.edu}

\flushbottom \maketitle

\begin{abstract}
{A laboratory experiment is suggested in which conditions similar to those in the plume ejecta from Enceladus and, perhaps, Europa are established.  Using infrared spectroscopy and polarimetry, the experiment might identify possible bio-markers in differential measurements of water from the   open-ocean, from
hydrothermal vents, and abiotic water samples. 
Should the experiment succeed, large telescopes could be used to acquire sensitive infrared spectra of the plumes of Enceladus and Europa, as the satellites transit the bright planetary disks.   The extreme technical challenges encountered in so doing are similar to those of solar imaging spectropolarimetry. The desired signals are buried in noisy data in the presence of seeing-induced image motion and a changing natural source. Some differential measurements used for solar spectropolarimetry can achieve 
S/N ratios of $10^5$ even in the presence of systematic errors two orders of magnitude larger.}   We
review the techniques and likelihood of success of such an observing
campaign with some of the world's largest ground-based telescopes, as
well as the long anticipated James Webb Space Telescope.    We discuss the relative merits of the new 4m Daniel K. Inouye Solar Telescope, as well as the James Webb Space Telescope and larger ground-based observatories, for observing the satellites of giant planets.  
As seen from near Earth, transits of Europa occur
regularly, but transits of Enceladus will begin again only in 2022. 
\end{abstract}

\section*{Introduction}

In humanity's perennial search for extraterrestrial life, one object
seems particularly promising:

\begin{quote}
``Enceladus has\ldots a textbook-like list of those properties needed
  for life\ldots [and] the ultimate free lunch: jets that spurt
  organic material into space" -- Catling (2013) \cite{Catling2013}
\end{quote}
 
The remarkable story of discoveries about Enceladus by the Cassini
Mission and science teams can be found in \cite{Spencer+Nimmo2013},
with post-2013 updates at a JPL webpage\footnote{\small
https://saturn.jpl.nasa.gov/news/2916/\-cassini-at-enceladus\--a\--decade\--plus\--of\--discovery/}.
Several lines of evidence, including {\em in-situ} sampling
of the ejecta as well as imaging and spectral data, indicate that the
plumes contain material similar to that found in hydrothermal vents in
Earth's deep oceans (e.g., \cite{Hsu+others2015}).  To produce properties of some of 
the ejected rock grains from Enceladus, the water temperature would somewhere have to exceed
$\approx 90^\circ$C.   Cassini gas phase CH$_4$/hydrocarbon
abundance ratios $\sim 10^2$ are compatible with abiotic sources. But
these measurements do not reject some production by micro-organisms found on Earth called methanogens   (e.g., \cite{Catling2013}), 
which, as extremophile organisms in hydrothermal vent environments on Earth, produce CH$_4$/hydrocarbon
abundance ratios $\geq 10^3$. This is simply because too little is known about the chemical history of Enceladus, and we know nothing about
possible biochemical environments there. 
Europa has three
reported episodes of emission of plumes from its interior
\cite{Roth+others2014,Sparks+others2016ApJ,Sparks+others2017}. Its plume emissions seem to be rare compared with Enceladus.

\re{In the last four decades, research on
hydrothermal vent
environments has revealed diverse and abundant life forms, living primarily on heat and chemistry.  
The possible importance of such colonies of 
non-photosynthetic life for originating all life on Earth has
been widely discussed
(e.g., \cite{Gold1992,Gold1999,Catling2013}). The hydrothermal vents are distributed along the Earth's tectonic plate boundaries. \re{Tectonic activity is} 
frequently listed as a prerequisite for habitability of planets, \re{continually bringing mineral-rich material to the
surface}.   Some structures on the S. polar surface of Enceladus have been described as ``tectonic'' \cite{Spencer+Nimmo2013}.

Two classes of vents host very different 
ecosystems.  Most relevant to this paper are the old (at least 30,000 years), alkaline,  90 C vents, typified by the ``Lost City Hydrothermal Field'' (LCHF)
\cite{Brazelton+others2006}. 
The vents efficiently release CH$_4$ and H$_2$, unlike their  hotter (350 C), acidic ``black smoker'', 100$\times$  younger counterparts which produce CO$_2$, H$_2$S and some metals.   LCHF and black smoker vents support different lifeforms.
The LCHF ecosystems are believed appropriate to the Jovian and Saturnian 
satellites, 
but at this stage one should
not reject out-of-hand the possible importance of the black smokers.  In the black smoker ecosystems, microbial organism concentrations 
are some $10^4$ to $10^4$ higher than non-venting regions.
The LCHF contains of order $10^5$ cells cm$^{-3}$ in the LCHF
\cite{Brazelton+others2006,Porco+others2017}, compared with the $3\times10^{21}$ number density of water molecules.  We cannot expect to detect directly such cells, but the number densities of much smaller  biogenic molecules associated with 
such cellular life should be much larger. 
}

It 
seems important to try to detect signs of life in the material ejected
from both Enceladus and  Europa by whatever means possible.
Unfortunately, the earliest planned fly-by and lander will not even launch before 2022, even for Jupiter, pushing back encounters until
after $2028$. This paper addresses the question, \textit{might
  we probe this organic material remotely, and attempt to provide the
  first evidence for extra-terrestrial, simple life?}  We will conclude,
surprisingly, that we already have in place both the needed
instrumentation and techniques to attempt such measurements. However,
new laboratory work is also needed to mimic conditions of the water
ejected by the satellites into space.  \re{So here we} put forward a
program of research involving extremely high sensitivity
imaging spectroscopy, routinely used in solar
work, together with some of the most advanced telescope systems on the
ground and in space, to attempt such measurements. 

\re{Specifically, t}he ideas
expounded here \cite{Judge2016} are \textit{to measure, differentially, the
  absorption spectrum of the plumes as each satellite transits the
  parent disk}.  
Such measurements will record dips in the planetary
light as both the opaque satellite and the plume material make their
disk passages. 
\re{Circumstances of the transits are listed in Table~\protect\ref{tab:circ}.}

\tabone

\re{To proceed, we first look at similarities 
between the transit spectroscopy and solar spectro-polarimetry.  Then we propose laboratory work to see if spectral bio-signatures exist in water sampled from diverse biological habitats in the Earth's oceans. Lastly, we explore the feasibility of the proposed  research using some of the world's largest telescopes.}

\section*{Commonalities with solar spectropolarimetry}

\re{These ideas have a superficial similarity to work exploring exo-planetary atmospheres, both use transits and both seek weak signals against a very bright background. But there are significant technical differences: Firstly, exoplanet transits are spatially unresolved, satellite transits must be spatially resolved in order to fill as much of each pixel with 
plumes;
secondly, satellite transits are subject to detrimental seeing-induced noise as images are blurred rapidly in time by Earth's atmosphere; lastly, 
 as the satellite/plume advances across the planetary disk, the background scene is changing in time.

All-in-all, the proposed observations of satellite transits have much more in common with solar work, in particular solar {\em spectro-polarimetry}, than with the exoplanet transit work. Modern solar observations at visible and infrared wavelengths are
generally performed near the diffraction limit using adaptive
optics, image reconstruction techniques, and splitting  light into both
wavelength and polarization states.  
Several authors (e.g., \cite{Landi2013, Judge2017cjp}) have
recently reviewed the challenges facing modern solar
spectropolarimetry.  

The commonalities in the needs for transit spectroscopy and solar spectro-polarimetry are as follows:

\begin{itemize}
\item Both require very high signal-to-noise ratios.  In the solar case, information on the magnetic field is often encoded in signals as small as $10^{-4}$ of
the measured intensity, in the plumes, the small optical depths and geometric sizes of plume material will lead to similarly small signals of interest.

\item The highest angular resolutions possible, close to
  diffraction limits, are needed in both cases. In the Sun, we try to resolve spatially intermittent magnetic field interacting with plasma at the smallest scales possible, and Enceladus' plumes 
are a mere 
  0.016'' long, filling a small fraction of the area of the spectrograph slit.

\item The small physical scales and rapid changes of the Sun's
  magnetic field, and of the plumes and their transit across planetary
  features, both set limits on the largest acceptable exposure times  (Table~\ref{tab:circ}).
\item Rapid ($\gg 1 $Hz) 
variations in the seeing conditions. presents a serious
  problem.  Adaptive optics (AO) must be brought to bear
 because the targets (e.g., sunspots on the Sun, satellites on the planet's disk) show structured objects covering small angular areas.
\end{itemize}
}
One advantage presented by  satellite plume observations is that, unlike the Sun, we can simply sum all exposures, because we  \re{seek an average spectrum}.  In contrast, modern solar data require integration times of at most seconds to avoid smearing dynamical phenomena of interest. This difference makes up, to some degree, for the much dimmer planetary surfaces. 

\section*{Needed laboratory work}

Transmission spectra of
seawater should be obtained in the laboratory, between the atmospheric
cutoff at 390 nm and, say 10 $\mu$m, \re{ideally} with a resolution
$\geq 10^4$. 
High intensity infrared and visible light sources can be used to obtain transmission spectra through the expanded vapor.  

 To approach the very low density and pressure conditions at the plumes in space, a sample of liquid water  \re{ might be made to }expand into a vacuum.   The number density of water molecules in the plumes can be estimated using the scale lengths of
Table~\ref{tab:circ} and measured column densities of
$1.5\times10^{16}$ cm$^{-2}$ \cite{Hansen+others2006}.  The observed
columns \cite{Hansen+others2006} are through the jet-like structures,
which are of order a factor $10$ smaller than the scale lengths in
Table~\ref{tab:circ}.  Thus, with a path length of around 10 km for
Enceladus, we find an average molecular density of water of $n_{H_2O}
\approx 1.5 \times 10^{10}$ molecules/ cm$^3$.

\re{By most laboratory standards, this is a very low density.   } 
Using a sample of say,
0.1 cm$^3$ of liquid water, which contains $\approx 6\times10^{23}
\ \times 0.1\ / \ 18\approx 3\times10^{21}$ molecules, densities
inside a vacuum chamber of volume ${\cal V}$ cm$^{3}$ are $\approx
3\times10^{21}/{\cal V}$.  To produce densities close to those of the
plumes would require ${\cal V} \approx 2\times10^{11}$ or a vacuum
chamber of size $L \approx 59 $ meters, with a characteristic path
length of only $Ln \approx 10^{14}$ molecules cm$^{-2}$.  Instead we
consider a vacuum chamber of linear size 1 meter or so, yielding an expanded density $n
\approx 3 \times 10^{15}$ cm$^{-3}$ and a \re{column density} $nL \approx 3\times10^{17}$
cm$^{-2}$.  The latter, which should be high 
to produce measurable absorption spectra, 
 can be increased by allowing water to expand
into an oblate vacuum chamber for the same volume, by factors of the
aspect ratio (length/width).   \re{By way of comparison, 
Table~\ref{tab:circ} lists column density an order of magnitude lower for the Enceladus plumes, which is remarkably close for such diverse conditions. 
However, post-expansion number densities of H$_2$O in the vacuuum chamber  $n\approx 3\times10^{15}$
cm$^{-3}$ are 5 orders of magnitude larger than in the plumes, with  
mean inter-molecular distances of $\Lambda
\approx 7\times10^{-6}$ cm compared with $4\times10^{-4}$ cm for
conditions in Enceladus' plumes}.  Amino acids
contain upwards of a few hundred atoms, each of size $10^{-8}$ cm,
small bacteria are $\geq 10^{-5}$ cm across, prokaryotes $\approx 3
\times10^{-4} $ cm.  In both the plumes and laboratory vessel, any large (biological) molecules or even organisms will be
embedded in a very cold and tenuous H$_2$O vapor, likely with ice
particles.   

\re{Assuming adiabatic expansion with an exponent of 5/3, 
the final pressure would approach 0.003 dyne~cm$^{-2}$ (a ``high vacuum'' at $p=3\times10^{-9}$ atmospheres) probably
requiring multi-stage pumping with an ion-gauge measurement. 
The temperature of the vapor in the experiment would be $0.006$ K, the mean speed 
of H$_2$O molecules $170$ cm~s$^{-1}$. With a gas-kinetic cross section of order $10^{-15}$ cm$^2$, we find a collisional mean free path of 0.3 cm and a collision time of
0.002 sec.  For isothermal expansion,  
the collision time would be reduced to $10^{-5}$ sec and the pressure increased to $10^{-4}$ atmospheres.  The isothermal and adiabatic approximations represent the limits of short and long energy exchange times respectively. In both cases the 
transmission spectra should be similar since we will be far from sampling optically thick material. }
The higher \re{number} density $\approx 3\times 10^{15}$ cm$^{-3}$ of laboratory
vapor can lead to changes in the spectra of large molecules via ``collisions'', because the
molecules are packed a factor of 50 closer in the laboratory than in
plumes.  It is appropriate therefore to vary the densities of
the water molecules to look for systematic effects
of collisions between any larger molecules and the water vapor substrate. 

What kind of water ecosystems should be measured?  Known
oceanic ecosystems on Earth are based on \re{only two sources of energy
(e.g., \cite{McKay+others2008}):
sunlight and chemical energy, 
the second of which was recognized
only in 1979 \cite{Corliss+others1979}.
In the absence of sunlight, deep in the
ocean there is abundant life deriving its energy from {\em chemosynthesis}. 

The purpose of the laboratory work is therefore 
to see if such bio-signatures appear detectable through spectroscopy, for we cannot judge from existing work what signatures might be present. 
We anticipate performing an experiment along the following lines.  A samples of various sources  of sea- and fresh-water
should be measured differentially against one another. These must include

\begin{enumerate}
\item Water from several regions close to hydrothermal vents that are abundant in chemosynthetic life forms, from both high temperature (300 C) acidic 
(black smoker) and 
low temperature (90 C) alkaline fumaroles should be examined,
\item Normal seawater,
\item Water from land-surface geysers.
\end{enumerate}
}
In all cases differential measurements of the same water samples, but with large molecules and organisms removed (by physical and/or chemical means), should be made. 
The experiment might proceed as follows.  A vacuum chamber equipped with a
suitable window, and with a volume of $\approx1$ m$^3$
should be dried and pumped down to less than \re{$10^{-9}$ atmospheres}.  A cell
containing water samples can be 
suddenly opened to the vacuum chamber.  \re{During this dynamical
expansion and relaxation phase, time-resolved spectra should be obtained, using a grating 
IR spectrometer owing to the small dynamical times of less than a second. }
Spectra
should be obtained prior to and after the rapid expansion phase to allow
differential measurements.  The experiment can be repeated until
sufficient S/N ratios are achieved ($\geq 10^4$).  Spectral
sensitivity to different pressures and temperatures should be investigated.   Finally, it might be that, with the helical handedness of
many biomolecules, attempts at circular polarization measurements
might be profitable. It is not  inconceivable that such measurements
will ultimately help us understand the overwhelming bias of life on
Earth towards one chirality.

\figboth

\section*{Calculations}

Here we will show that the transit measurements proposed are feasible. The question
of whether or not {\em bio}-signatures can be detected depends critically \re{on the outcome of the laboratory work, and on acquisition of the highest possible signal-to-noise ratios of the planets.   We proceed in a spirit of scientific exploration of the unknown, assuming that the laboratory work is successful. } 
Here we perform some order-of-magnitude
calculations to assess the likelihood of success.  We will present
Enceladus in detail, showing the experiments to be worthwhile. The numbers in Tables \ref{tab:circ} and \ref{tab:obs} show that Europa is a far easier target, if it can be caught during a rare episode of ejection of matter. 

Figure~\ref{fig:both} shows the geometry
of transits for the two satellites, together with boxes that represent spatial pixels
of angular size 0.05" and 0.1" that are representative of \re{conditions} under which  observations appear possible (cf. \ref{tab:obs}).
\re{A balance must be struck between angular resolution and the need to detect plume absorption. 
While absorption cross sections are high at UV wavelengths, and diffraction-limited angular resolution is also 
high, UV photon fluxes are very low.  }
Photon fluxes from scattered sunlight from the planetary atmospheres are 3 and 2 orders of magnitude lower at 0.15 and 0.2 $\mu$m compared with 2 $\mu$m respectively (using reflectivities from   \cite{Morrissey+others1995}). 
Estimates (below) of optical depths of \re{even abundant species (such as CH$_4$) in the plumes 
show that they will be small, $\tau \approx 10^{-3}$}. The signals
desired will be a fraction $\approx \tau$ of the intensity.  \re{These signals will be diluted further because of the  small apparent sizes of the
plumes which, for 
Enceladus, lie below the resolution of most instruments,} and for Enceladus and Europa 
the detrimental effects of atmospheric seeing must be
mitigated.  A further difficulty for Enceladus is that the velocity of
the transiting satellite limits integrations times to, at most, a few
seconds (see Table~\ref{tab:circ}), after which the plume intercepts a different
part of the planetary surface, and hence surface features, behind it.  \re{All of these considerations point to optimal 
wavelengths between 0.5 and 5$\mu$m, to find a balance between brightness (which falls rapidly at shorter wavelengths) and angular 
resolution (diffraction reducing the resolution at longer wavelengths, as $0.24'' \lambda[\mu{\rm m}]/D [{\rm m}]$, $D$=telescope diameter in m).}  In the calculations
below, we will see that, even at the brightest parts of the spectrum
of Saturn and Jupiter, we will be limited by photon noise. Observing between 0.5 and 5 $\mu$m means that
we will be probing signatures of vibration-rotation modes of large molecules.  \re{This region contains in principle a variety of 
spectral bio-signatures (e.g., \cite{Hand+others2009}).}

\figspec

\subsection*{Signal-to-noise estimates of transmission spectra for Enceladus}

First we compute the intensity (brightness) of Saturn between 0.5 and
5 microns, wavelengths at which instruments will operate
in space (the James Webb Space Telescope- ``JWST") and on the ground
(including DKIST).  Several well-known atmospheric transmission
windows (the VRI and J-M astronomical bands) allow measurements of
astronomical objects from the ground at these wavelengths.  Assuming for simplicity 
that light scattered from Saturn's cloud decks is uniformly emitted
outwards into $2\pi$ steradians, the reflected (scattered) light
intensity from the planetary surface is
\begin{equation}
I_\nu = \frac{f_\nu}{2\pi} a_\nu \ \ \ \intu
\end{equation}
where $a_\nu$ is the planetary albedo at frequency $\nu$, $f_\nu$ is
the flux density of solar radiation at Saturn (\saturn),
\begin{equation}
f_\nu \approx \frac{\pi R_\odot^2}{\Delta_{\saturn}^2} B_\nu(T_\odot)
\ \ \ \fluxau
\end{equation}
and where $B_\nu(T_\odot)$ is the brightness of the solar disk (here
given as a Planck function at $T_\odot\approx$5000K). 
$\Delta_{\saturn}$ is
the distance from the Sun to Saturn. 
\re{Then, retaining the notation 
$a_\nu B_\nu(T_\odot)$  no matter if the emission is scattering or thermal, recognizing that 
$a_\nu$ is given by the ratio of Saturn's emission spectrum to the Planck function curve in Figure~\protect\ref{fig:spec}, we have}
\begin{equation}
I_\nu \approx \frac{R_\odot^2}{2\Delta_{\saturn}^2} a_\nu B_\nu(T_\odot)
\ \ \ \intu
\end{equation}
The flux density from an area on the planet subtending a solid angle
$\omega$ steradians at a telescope near Earth is simply $\omega I_\nu$
\fluxau.  The flux density of photons is just $\omega I_\nu /h\nu$
\photau, so that for a telescope with an aperture of diameter ${\cal
  D}$ cm$^2$, and a total (telescope plus feed optics, spectrograph
and detector) system efficiency of photon detection of ${\cal E}$, we
find a photon counting rate $N_\nu$ (photons Hz$^{-1}$ s$^{-1}$) of
\begin{equation}
N_\nu = \frac{\pi}{8h\nu} \frac{R_\odot^2}{\Delta_{\saturn}^2} a_\nu
B_\nu(T_\odot) \ {\cal D}^2 {\cal E} \omega \ \ \ {\rm
  photons\ Hz^{-1}\ s^{-1}}
\end{equation}

It is clear because of the small size of plumes and their large distance that we must make observations close to the diffraction
limit of visible and infrared telescopes (Tables~\ref{tab:circ} and \ref{tab:obs}). At the diffraction
limit the angular size is close to $\vartheta \approx 1.2 \lambda /
{\cal D} $ radians, $\lambda=c/\nu$.  If we critically sample the Airy
disk using square pixels \re{at the telescope focus where the entrance slit to the spectrograph is placed}, we need
\begin{equation}
\vartheta_C \approx \frac{0.6 c}{\nu {\cal D}} \ \ \ {\rm radians.}
\end{equation}
For a ${\cal D}=4$ meter telescope observing at 4 $\mu$m, $\vartheta_C
= 0.12$ arcseconds.  Then
\begin{equation}
\omega \approx \vartheta_C^2 = 0.4 \left ( \frac {c}{\nu \cal D}\right
)^2
\end{equation}
so that the photon counting rate becomes simply
\begin{equation}
N_\nu = \frac{\pi}{4}\ \frac{R_\odot^2} {\Delta_{\saturn}^2} \ a_\nu
\ \frac{c^2}{2h\nu^3} B_\nu(T_\odot) \ {\cal E} \ \ \ {\rm
  photons\ Hz^{-1}\ s^{-1} \ Px}^{-1}
\end{equation}
independent of the telescope aperture, where 
Px refers to each spatial pixel.  Substituting for $B_\nu$ in the Rayleigh-Jeans limit we
find
\begin{equation}
N_\nu \approx \frac{\pi}{4} \ \frac{R_\odot^2}{\Delta_{\saturn}^2}
\ a_\nu \frac{kT_\odot}{h\nu}
%\left ( \frac{1}{\exp(h\nu/kT_\odot)-1} \right )
\ {\cal E}
\ \ \ {\rm photons\ Hz^{-1}\ s^{-1} \ Px}^{-1}
\end{equation}
For Saturn, at opposition $\Delta_{\saturn}=9.53$ A.U. =
$1.43\times10^{14}$ cm.  With $R_\odot = 6.996\times10^{10}$ cm, the
numerical values at $\lambda=4$ $\mu$m are
\begin{equation}
N_\nu \approx 1.8\times10^{-7} \ \ a_\nu \ {\cal E} \ \ \ {\rm
  photons\ Hz^{-1}\ s^{-1} \ Px}^{-1}
\end{equation}
For a spectrograph observing Saturn with resolution $\nu/\Delta\nu={\cal R}$, and
critically sampling in wavelength using a detector with spectral pixels Sx with
width $0.6 c /\nu{\cal R}$, we find, for $1 < \lambda < 5$ $\mu$m,
\begin{equation} \label{eq:n}
N_{Sx} \approx 4000 \left [ \frac{\lambda}{4 \mu {\rm m}} \right ]
\left[ \frac{3000} {{\cal R}} \right ] \ a_\nu \ {\cal E} \ \ \ {\rm
  photons\  s^{-1} Px^{-1} Sx^{-1} }.
\end{equation}
For Jupiter (\jupiter), the numerical constant of 4000 is simply
$\Delta^2_{\saturn}/\Delta^2_{\jupiter} = 3.4$ times higher. It must also
be remembered that Jupiter is closer to Earth (\earth) so that
potentially any plumes on Europa are far easier to resolve than on
Enceladus, for a given $\vartheta_C$ and $\omega$ (see
Figure~\ref{fig:both}).  The above equation allows us to
estimate the number of photons per second that can be used for AO
correction, using Saturn and the satellite as the source for the AO
corrections, noting that $\int N_\nu d\nu \approx 10^{15} N_\nu \geq
10^6$ photons s$^{-1}$ over the broad visible spectral range.  This
will give $\geq 10^3$ photons per spatial pixel,  when the AO bandwidth is 1
kHz.

\re{As stated above, the values of $a_\nu$  are simply the ratio of the plotted spectra from \cite{Clark+McCord1979}
and the ISO spectrum shown in Figure~\protect\ref{fig:spec} to the scaled Planck function to the dashed line, which varies as $\lambda^{-2}$ at wavelengths longer than those plotted. The brightest ``windows'' of emission in Figure~\protect\ref{fig:spec}  (broad peaks in the spectrum) all have $a_\nu \approx 0.3$, and this value is adopted below, recognizing that other regions of the spectrum will be considerably dimmer.}

While the upper panel of Figure~\ref{fig:spec} represents the
background source against which we might attempt to measure the
transmission spectrum of the plumes of Enceladus, the lower panels
show calculations of the expected transmission of light through the
plumes.  These calculations include just the abundant molecules found
in mass spectrometry work by \cite{Waite+others2006}: H$_2$O, CH$_4$,
CO$_2$, O$_2$. All molecules were assumed to be in the gas
phase. \cite{Hansen+others2006} showed that Enceladus's plumes are at
least partly in the gas phase.  We adopt the relative abundances of
\cite{Waite+others2006}, H$_2$O (91\%), CH$_4$ (1.6\%), CO$_2$ (3\%),
O$_2$ ($<1$\%).  The H$_2$O molecular column density was set to $1.5\times
10^{16}$ cm$^{-2}$, determined from transmission spectra of the UV
bright star $\gamma$ Orionis during a flyby of Cassini in 2005, and
the plume path length was set to the scale height of the observed
plumes, $\sim 10^2$ km \cite{Hansen+others2006}.  The computed absorption
depths of molecular lines are, as expected, roughly in proportion to
the molecular abundances.  We emphasize several features of
Figures~\ref{fig:spec}.  Firstly, the dominant absorbers leave plenty
of spectral ``room'' for detection of other molecular species.
Secondly, the emission spectrum from Saturn, while spectrally highly
structured (Figure~\ref{fig:spec}), offers a bright background \re{($>10$ Jy)} except
for the gap between 3.4 and 4 $\mu$m.  Thirdly, we see that many lines
have absorption depths less than 0.001, even though these molecules
have relative abundances by number exceeding 1\%.  In order to perform
the proposed experiments it is clear that {\em we must achieve the
  highest possible signal-to-noise ratios. Any experiment should try
  to achieve a sensitivity of better than $10^{-4}$ of the brightness of
  the background spectrum of Saturn. } This criterion implies
acquiring at least $10^8$ photons per spectral range of interest (it
could be one spectral pixel or many pixels that all correspond to
features discovered in the spectra of water samples on Earth, discussed below).

\subsection*{How to achieve the required signal-to-noise ratios}

The transit durations are several hours (Table \ref{tab:circ}).  \re{Using
$a_\nu \approx 0.3$, a system efficiency ${\cal E} \approx 0.3$, we have 400 photons per
spectral pixel Sx per spatial pixel Px per second.  This applies to an imaging
system critically sampling the diffraction limit,
something that is undesirable in solar work 
owing to limited exposure times on the same solar 
scene \cite{Landi2013,Judge2017cjp}, but which is
not a problem here as it is only the background scene that is varying during orbital motion. } In one transit, 
this system will accumulate $3\times10^6$ photons \re{per Sx and per Px}.  Given the very small angular sizes of the target plumes, we must avoid binning spatially.  We might bin $n_s$ Sx pixels,
then we would acquire $3\times10^6 n_s$ photons per spectral
region of interest per transit. \re{Thus one transit will require
$n_s > 36$ to acquire $10^8$ photons per spectral element. 
This can be achieved with a spectral resolution ${\cal R } \approx
80$, for example.  By observing 10 consecutive transits one could
accumulate $10^9$ photons under the same telescope/instrument
configuration.  The success or failure of this spectral
measurement can then be seen to depend critically on the 
presence of broad features in the samples from the laboratory spectrum.

Thus, photon counting statistics limit the achievable
signal-to-noise ratios to the extent that a spectral resolution of 80 appears insufficient, which can only be determined by performing the laboratory experiment. } It is likely that systematic errors
induced through residual image motions, inaccurate flat-fields and
dark currents, instrumental secular changes in sensitivity and other
instrumental factors will, uncorrected, limit a set of measurements to
far larger systematic noise errors.
This is where experience in observational solar physics can help, for
ground-based solar data are plagued with similar issues.  One of the
major problems involves intrinsic and seeing-induced image motion of
bright, extended objects, which introduces spurious time-dependent
signals from neighboring pixels into the data.  Such problems are
absent from unresolved sources such as stars, which with care can
achieve sensitivites of $10^5$ by deep integrations and co-addition of
many spectral lines \cite{Bagnulo+others2009}. Yet signal-to-noise
ratios on the order of $5\times 10^{3}$ can be routinely obtained for
the Sun \cite{Collados+others2007}, sometimes approaching $10^5$
\cite{Gandorfer+others2004}, even in the presence of rapid image
motions.  These sensitivities are achieved using a combination of all
or some of the following: (1) differential techniques, including split
optical beams, beam switching; (2) rapid data acquisition; (3)
adaptive optics.

\figpos

Figure~\ref{fig:pos} shows an example of how differential measurements
might achieve the needed signal-to-noise for the case of transits of
Enceladus. While Saturn is over 200 times the diameter of Enceladus,
modern telescope systems with AO can correct seeing-influenced images
down to rms errors of around 30 mas (the unfilled circle shows a 30
mas radius superposed on the image). The dashed boxes show a $1\sigma$
excursion of seeing-induced motions corrected by a good AO system.
During an exposure of the spectrograph of order 1-10 seconds, light
will enter each of the ``pixels" shown from a random distribution of
such excursions.  Now, let us consider how we might attempt to reach
the highest s/n ratios with such measurements.

We wish to recover the absorption spectra of
the S. polar plumes which occupy a small
area of pixels in Figure~\ref{fig:pos}.  We will assume that plumes are present during
the entire duration of the transit.  Now, pixel $n-1$ has already been
exposed to Saturn's light through the plumes, some $\Delta t \approx
25$ seconds or so earlier than the image shows, for pixels of size
0.05 arcsecsonds.  Pixel $n$ is, at the time shown, exposed to the
plumes, and pixel $n+1$ has yet to be exposed to the plumes.  The time
scale of 25 seconds is important for several reasons. On this time
scale, we can assume that the underlying light emission by Saturn
remains constant, it is modified only by Saturn's rotation of
its cloud belts at the latitude observed.  Close to the equator,
Saturn's rotation period is about 10 hours and 14 minutes. Close to
the center of Saturn's disk the cloud decks rotate at roughly 1.6
\velu{}, almost 8 times slower than the orbital velocity of Enceladus
across the disk, corresponding to $\approx 2.6\times10^{-4}$
arcseconds per second, relative to the system's barycenter when the
system is at opposition.  For simplicity of exposition here, let us
treat Saturn as unchanging during exposures of order 25 seconds or so.
(Of course, such corrections will be applied in any final
analysis). Then, for each spectral pixel, assuming Saturn's brightness
itself is unchanging, and the instrument is stable, we find that the
counts $C_{n\ell }(\tau)$ in spatial pixel $n$ for each spectral pixel
$\ell$ at a time with index $\tau$ is given by
\begin{equation}
C_{n\ell }(\tau) = g_{n\ell} I_{n\ell }(\tau) + d_{n\ell},
\end{equation}
where $I_{n\ell }(\tau)$ is the intensity imaged on to spectral pixel
$\ell$ at spatial pixel $n$, averaged over the exposure time centered
at time index $\tau$.  $I_{n\ell }(\tau)$ includes all of the unknown
(except in a statistical sense) seeing-induced and/or instrumental
jitter image motions.  The detector plus system's gain is given by
$g_{n\ell} $, independent of time index $\tau$ (otherwise the detector
is a very poor one), and similarly $d_{n\ell}$ is the dark current
correction.  Now some 20 seconds later, the counts at time index
$\tau+1$ are
\begin{equation}
C_{n\ell}( \tau+1) = g_{n\ell} I_{n\ell}( \tau+1) + d_{n\ell}.
\end{equation}
Subtracting dark currents and dividing these two equations we obtain
the ratio of the plume intensity to the non-plume intensity, {\em for
  the same region of the planet} simply as follows:
\begin{equation} \label{eq:final}
\frac{I_{plume}}{I_{non-plume}} = \frac{I_{n\ell }(\tau)} {I_{n\ell
    \tau+1}} = \frac { C_{n\ell }(\tau) -d_{n\ell} } {C_{n\ell}( \tau+1) -
  d_{n\ell} },
\end{equation} 
independent of the gains of each spectral pixel. This manipulation is
a trick similar to that used to obtain very high signal-to-noise
ratios in stellar spectropolarimetry \cite{Bagnulo+others2009}, to
avoid dealing with gain corrections. With $\approx 450$ such
differential measurements for a full transit, we get as before with
${\cal R} = 3000$, $3\times10^6$ photons per spectral and spatial
pixel, for both $C_{n\ell }(\tau)$ and $C_{n\ell }(\tau+1)$ respectively.
In this case the s/n ratio due to photon statistics would be, assuming
changes in dark current are negligible (i.e. using a good detector
with $C_{n\ell}(\tau) \gg d_{n\ell}$ and $d_{n\ell}=$ constant over a
few minute period), $\sqrt[]{2}$ times higher than the noise at one
time $\tau$ (through the propagation of errors in both $C_{n\ell
  }(\tau)$ and $C_{n\ell}( \tau+1)$).  However, this factor can be reduced
to near unity by suitably averaging data for $C_{n\ell \tau+m}$ for
$m=-10$ to $m=+10$ say on the denominator of
equation~(\ref{eq:final}), with the assumption that observing
conditions do not change in the period of $20$ times 20 seconds, a few
minutes.  Finally, the spectrum desired $I_\ell$ can be obtained by averaging over all the best exposures.

It seems clear that, sacrificing spectral resolution, and assuming
that AO can produce imaging quality with rms seeing of around 30
milli-arc-seconds, the differential measurements represented by the
scheme shown in Figure \ref{fig:pos} and in equation (\ref{eq:final})
can get close ($\approx 2000$) to the desired s/n ratios ($\geq 10^4$)
for Enceladus, for just one transit. These techniques are standard in
both solar and stellar spectropolarimetry.  It should be noted that
Enceladus is especially challenging owing to its distance, and
relatively small size, which means that modern telescopes cannot
resolve the ``plumes''. The plume spectra are therefore diluted further
by the ratio of the fractional areas of the plume material in each
pixel (see Figure~\ref{fig:pos} for a general idea).  In every
technical sense, Europa is a far easier target: the surface brightness
of Jupiter is larger, the documented plumes are higher, and Europa regularly transits Jupiter's disk.  
\re{Signal-to-noise ratios for Jupiter and Europa are 
larger by a factor of 3.4 (equation~\ref{eq:n}) and  another order of magnitude because the Europa plumes should fill far more of each spatial pixel.}
Yet its eruptive events appear rare, they are 
less-well documented. Catling's ``free lunch'' \cite{Catling2013} has its limits.  

\subsection*{A comparison of  observatories}

In Table~\ref{tab:obs} we compare relevant IR  capabilities of three
observatories.  Both the JWST and DKIST telescopes are under
construction, while Keck telescopes have been in operation since 1993.
DKIST is included because, being primarily a solar telescope, it is
likely to have less pressure for night-time observations, and because
it has interesting capabilities.  In particular, the adaptive optics
system is designed to vary on a resolved bright source, not on point 
sources, and it is designed to do full Stokes polarimetry.   Enceladus has one of the brightest surfaces in the solar system, and will likely be brighter than Saturn's disk at the wavelengths considered. One  disadvantage of DKIST is the relatively high spectral dispersion of the first-light instrument CRYO-NIRSP, which reduces photon fluxes per pixel.  
But on the other hand, it is also a coronagraph, which makes it attractive for different kinds of observations of giant planet moons.  
For example, (see e.g.,  section 4.2 of \cite{Keszthelyi+others2016}) note the need for observations with low stray light while certain moons enter the shadow of their host planet, always very close to the planet itself as seen from Earth. 

Let us first consider the ground-based observatories.  Referring again to
Figure~\ref{fig:pos} and Table~\ref{tab:obs}, it is easy to see that the spectrum $I_{plume}$
will contain light from Enceladus' surface during each integration as
the residual seeing excursions move the sky image in and out of the
spectrograph pixels.  For observations from the ground, this
contribution must be corrected.  \re{Quantifying the contributions to noise is a (relatively) straightforward 
issue once the brightness gradients between the 
various objects in the seeing disk are quantified 
\cite{Lites1987,Judge+others2004}, and if the seeing power spectrum is available.  Calculations would need to be 
done if the laboratory experiment succeeds. One major advantage of the transit scenario instead of solar observations is that one can observe }
Enceladus directly
above the limb of the planet prior to and after transit to determine
the spectral nature of this contribution.  Clearly observations from
space, for example from the upcoming James Webb Space Telescope
(JWST), can remove seeing-induced contamination when the spacecraft
jitter is small enough. The JWST stability requirement ($< 3.7$ mas)
and NIRSpec focal plane geometric distortions ($<10$ mas)
\cite{Dorner+others2016} are sufficient to acquire high
quality plume spectra. However, JWST is not ideally suited to such observations, essentially because it was designed for  observing much fainter objects, and the pixels under-sample the diffraction limit at the shortest IR wavelengths.
This has two obvious consequences: (1) the pixel sizes of the instruments are larger than the plumes, and 
(2) the larger pixels collect more light, leading to saturation of the detectors 
at least for {\em imaging} of Jupiter and Saturn's disks.  By design, the saturation limits of the NIRSpec {\em spectrometer} on JWST, operating at its highest dispersion of ${\cal R} =2700$, shown in Figure~2 of \cite{Norwood+others2016}, lie above the count rates for the  expected brightness of all four gas giants in the solar system. 

\re{Other observatories have been examined in addition to these examples. The CRIRES spectrometer at the one of the VLT telescopes (${\cal D=}8.2$ m) has ${\cal R} = 10^5$ which is rather poorly matched to the much lower spectral resolution required to produce high count rates. The KMOS, NACO  and SINFONI instruments on the VLT seem as well suited as the Keck II instrument, the VLT has the MAD multi-conjugate adaptive optics system that has produced 0.09'' resolution images of Jupiter \footnote{https://www.eso.org/public/images/eso0833a/}. 
Coronagraphic instruments are less likely to be useful
since they introduce seeing-induced variations in brightness in targets such as bright transiting satellites, where the entire scene is bright. 
}

\tabobs

In conclusion, it seems that observatories exist, and will soon come
into operation, which can in principle investigate the transmission
spectra of plumes of Enceladus.  Any plumes detected again on Europa
would be far easier targets, should Europa emit additional plumes.

\section*{Conclusions}

This paper demonstrates the feasibility of making interesting 
measurements of plumes erupting from the surface of Enceladus, and
perhaps Europa. Astronomical and laboratory experiments can and should
be performed to try to detect signatures of biological products in the
transmission spectra during transits as Enceladus crosses the bright
disk of Saturn. The NIRSpec instrument on the JWST can obtain very
high quality differential spectra between 1 and 5 $\mu{}$m, but it has
rather large pixels which will dilute the signals of plume
material. Ground-based measurements will face the problem of dilution
of signals by residual seeing motions on scales larger than the plumes
of Enceladus.  The situation is different at Jupiter, where any plumes
present on Europa are of a much larger physical scale and easier to
detect spectroscopically. The problem is, of course, that Europa
clearly erupts rarely.

Lastly, since Enceladus' plumes supply Saturn's E ring with material,
then similar work when the E ring is close to being "edge-on" but
visibly separate from the more massive rings would seem worthwhile.
The polarization and perhaps coronagraphic credentials of DKIST might be used to advantage in such observations, as well as observations of giant planet satellites that are in the host planet's shadow.  In situations where the desired target lies very close to the very bright planetary disk \cite{Keszthelyi+others2016}, coronography might be particularly valuable. 

\vskip12pt
\noindent I am grateful to Wenxian Li for her comments and interest in
the work presented here. The two anonymous referees greatly helped to
improve the paper, and the author thanks Carolyn Porco for her
thoughts and encouragement.

%\bibliography{sample}

\small{\bibliographystyle{unsrt} \bibliography{main}

\begin{thebibliography}{10}

\bibitem{Catling2013}
D.~C. Catling.
\newblock {\em Astrobiology}.
\newblock Oxford, 1st edition, 2013.

\bibitem{Spencer+Nimmo2013}
J.~R. {Spencer} and F.~{Nimmo}.
\newblock {Enceladus: An Active Ice World in the Saturn System}.
\newblock {\em Annual Review of Earth and Planetary Sciences}, 41:693--717, May
  2013.

\bibitem{Hsu+others2015}
H.-W. {Hsu}, F.~{Postberg}, Y.~{Sekine}, T.~{Shibuya}, S.~{Kempf},
  M.~{Hor{\'a}nyi}, A.~{Juh{\'a}sz}, N.~{Altobelli}, K.~{Suzuki}, Y.~{Masaki},
  T.~{Kuwatani}, S.~{Tachibana}, S.-I. {Sirono}, G.~{Moragas-Klostermeyer}, and
  R.~{Srama}.
\newblock {Ongoing hydrothermal activities within Enceladus}.
\newblock {\em Nature}, 519:207--210, March 2015.

\bibitem{Roth+others2014}
L.~{Roth}, J.~{Saur}, K.~D. {Retherford}, D.~F. {Strobel}, P.~D. {Feldman},
  M.~A. {McGrath}, and F.~{Nimmo}.
\newblock {Transient Water Vapor at Europa's South Pole}.
\newblock {\em Science}, 343:171--174, January 2014.

\bibitem{Sparks+others2016ApJ}
W.~B. {Sparks}, K.~P. {Hand}, M.~A. {McGrath}, E.~{Bergeron}, M.~{Cracraft},
  and S.~E. {Deustua}.
\newblock {Probing for Evidence of Plumes on Europa with HST/STIS}.
\newblock {\em ApJ}, 829:121, October 2016.

\bibitem{Sparks+others2017}
W.~B. {Sparks}, B.~E. {Schmidt}, M.~A. {McGrath}, K.~P. {Hand}, J.~R.
  {Spencer}, M.~{Cracraft}, and S.~{E Deustua}.
\newblock {Active Cryovolcanism on Europa?}
\newblock {\em ApJL}, 839:L18, April 2017.

\bibitem{Gold1992}
T.~{Gold}.
\newblock {The Deep, Hot Biosphere}.
\newblock {\em Proceedings of the National Academy of Science}, 89:6045--6049,
  July 1992.

\bibitem{Gold1999}
T.~{Gold}.
\newblock {\em {The Deep Hot Biosphere}}.
\newblock 1999.

\bibitem{Brazelton+others2006}
W.~J. Brazelton, M.~O. Schrenk, D.~S. Kelley, and Baross~J. A.
\newblock Methane- and sulfur-metabolizing microbial communities dominate the
  lost city hydrothermal field ecosystem.
\newblock {\em {Applied and Environmental Biology}}, 72:6257–6270, September
  2006.

\bibitem{Porco+others2017}
C.C. Porco, L.~Dones, and C.~Mitchell.
\newblock Could it be snowing microbes on enceladus? {Assessing} conditions in
  its plume and implications for future missions.
\newblock {\em Astrobiology}, this volume, 2017.

\bibitem{Judge2016}
P.~G. Judge.
\newblock Unpublished poster paper: ``\protect{Bio}-signatures from
  \protect{Enceladus}' geysers using transits from 2023''.
\newblock In {\em Exploring the Universe with JWST II. Science Meeting October
  24 - 28, 2016. Montreal, Canada}, October 2016.

\bibitem{Landi2013}
E.~{Landi Degl'Innocenti}.
\newblock {Spectropolarimetry with new generation solar telecopes }.
\newblock {\em Memorie della Societa Astronomica Italiana}, 84:391, 2013.

\bibitem{Judge2017cjp}
P.~G. Judge.
\newblock Atomic physics and modern solar spectropolarimetry.
\newblock {\em Canadian J. Phys}, page in press, 2017.

\bibitem{Hansen+others2006}
C.~J. {Hansen}, L.~{Esposito}, A.~I.~F. {Stewart}, J.~{Colwell}, A.~{Hendrix},
  W.~{Pryor}, D.~{Shemansky}, and R.~{West}.
\newblock {Enceladus' Water Vapor Plume}.
\newblock {\em Science}, 311:1422--1425, March 2006.

\bibitem{McKay+others2008}
C.~P. {McKay}, {Porco Carolyn C.}, T.~{Altheide}, W.~L. {Davis}, and T.~A.
  {Kral}.
\newblock {The Possible Origin and Persistence of Life on Enceladus and
  Detection of Biomarkers in the Plume}.
\newblock {\em Astrobiology}, 8:909--919, October 2008.

\bibitem{Corliss+others1979}
J.~B. {Corliss}, J.~{Dymond}, L.~I. {Gordon}, J.~M. {Edmond}, R.~P. {von
  Herzen}, R.~D. {Ballard}, K.~{Green}, D.~{Williams}, A.~{Bainbridge},
  K.~{Crane}, and T.~H. {van Andel}.
\newblock {Submarine Thermal Springs on the Galapagos Rift}.
\newblock {\em Science}, 203:1073--1083, March 1979.

\bibitem{Morrissey+others1995}
P.~F. {Morrissey}, P.~D. {Feldman}, M.~A. {McGrath}, B.~C. {Wolven}, and H.~W.
  {Moos}.
\newblock {The Ultraviolet Reflectivity of Jupiter at 3.5 Angstrom Resolution
  from Astro-1 and Astro-2}.
\newblock {\em ApJL}, 454:L65, November 1995.

\bibitem{Hand+others2009}
K.~P. {Hand}, C.~F. {Chyba}, J.~C. {Priscu}, R.~W. {Carlson}, and K.~H.
  {Nealson}.
\newblock {\em {Astrobiology and the Potential for Life on Europa}}, page 589.
\newblock 2009.

\bibitem{Clark+McCord1979}
R.~N. {Clark} and T.~B. {McCord}.
\newblock {Jupiter and Saturn - Near-infrared spectral albedos}.
\newblock {\em Icarus}, 40:180--188, November 1979.

\bibitem{Rothman+others2013}
L.~S. {Rothman}, I.~E. {Gordon}, Y.~{Babikov}, A.~{Barbe}, D.~{Chris Benner},
  P.~F. {Bernath}, M.~{Birk}, L.~{Bizzocchi}, V.~{Boudon}, L.~R. {Brown},
  A.~{Campargue}, K.~{Chance}, E.~A. {Cohen}, L.~H. {Coudert}, V.~M. {Devi},
  B.~J. {Drouin}, A.~{Fayt}, J.-M. {Flaud}, R.~R. {Gamache}, J.~J. {Harrison},
  J.-M. {Hartmann}, C.~{Hill}, J.~T. {Hodges}, D.~{Jacquemart}, A.~{Jolly},
  J.~{Lamouroux}, R.~J. {Le Roy}, G.~{Li}, D.~A. {Long}, O.~M. {Lyulin}, C.~J.
  {Mackie}, S.~T. {Massie}, S.~{Mikhailenko}, H.~S.~P. {M{\"u}ller}, O.~V.
  {Naumenko}, A.~V. {Nikitin}, J.~{Orphal}, V.~{Perevalov}, A.~{Perrin}, E.~R.
  {Polovtseva}, C.~{Richard}, M.~A.~H. {Smith}, E.~{Starikova}, K.~{Sung},
  S.~{Tashkun}, J.~{Tennyson}, G.~C. {Toon}, V.~G. {Tyuterev}, and G.~{Wagner}.
\newblock {The HITRAN2012 molecular spectroscopic database}.
\newblock {\em JQSRT}, 130:4--50, November 2013.

\bibitem{Waite+others2006}
J.~H. {Waite}, M.~R. {Combi}, W.-H. {Ip}, T.~E. {Cravens}, R.~L. {McNutt},
  W.~{Kasprzak}, R.~{Yelle}, J.~{Luhmann}, H.~{Niemann}, D.~{Gell}, B.~{Magee},
  G.~{Fletcher}, J.~{Lunine}, and W.-L. {Tseng}.
\newblock {Cassini Ion and Neutral Mass Spectrometer: Enceladus Plume
  Composition and Structure}.
\newblock {\em Science}, 311:1419--1422, March 2006.

\bibitem{Bagnulo+others2009}
S.~{Bagnulo}, M.~{Landolfi}, J.~D. {Landstreet}, E.~{Landi
  Degl'Innocenti}, L.~{Fossati}, and M.~{Sterzik}.
\newblock {Stellar Spectropolarimetry with Retarder Waveplate and Beam Splitter
  Devices}.
\newblock {\em PASP}, 121:993, September 2009.

\bibitem{Collados+others2007}
M.~{Collados}, A.~{Lagg}, J.~J. {D{\'{\i}}az Garc{\'{\i}} A}, E.~{Hern{\'a}ndez
  Su{\'a}rez}, R.~{L{\'o}pez L{\'o}pez}, E.~{P{\'a}ez Ma{\~n}{\'a}}, and S.~K.
  {Solanki}.
\newblock {Tenerife Infrared Polarimeter II}.
\newblock In P.~{Heinzel}, I.~{Dorotovi{\v c}}, and R.~J. {Rutten}, editors,
  {\em The Physics of Chromospheric Plasmas}, volume 368 of {\em Astronomical
  Society of the Pacific Conference Series}, page 611, May 2007.

\bibitem{Gandorfer+others2004}
A.~M. {Gandorfer}, H.~P.~P.~P. {Steiner}, F.~{Aebersold}, U.~{Egger},
  A.~{Feller}, D.~{Gisler}, S.~{Hagenbuch}, and J.~O. {Stenflo}.
\newblock {Solar polarimetry in the near UV with the Zurich Imaging Polarimeter
  ZIMPOL II}.
\newblock {\em A\&A}, 422:703--708, August 2004.

\bibitem{Keszthelyi+others2016}
L.~{Keszthelyi}, W.~{Grundy}, J.~{Stansberry}, A.~{Sivaramakrishnan},
  D.~{Thatte}, M.~{Gudipati}, C.~{Tsang}, A.~{Greenbaum}, and C.~{McGruder}.
\newblock {Observing Outer Planet Satellites (Except Titan) with the James Webb
  Space Telescope: Science Justification and Observational Requirements}.
\newblock {\em PASP}, 128(1), January 2016.

\bibitem{Lites1987}
B.~W. Lites.
\newblock Rotating waveplates as polarization modulators for stokes polarimetry
  of the sun: evaluation of seeing-induced crosstalk errors.
\newblock {\em Applied Optics}, 26:3838--3845, 1987.

\bibitem{Judge+others2004}
P.~G. Judge, D.~F. Elmore, B.~W. Lites, C.~U. Keller, and T.~Rimmele.
\newblock Evaluation of seeing-induced cross-talk in tip/tilt corrected solar
  polarimetry.
\newblock {\em Applied Optics: optical technology and medical optics}, 43,
  issue 19:3817--3828, 2004.

\bibitem{Dorner+others2016}
B.~{Dorner}, G.~{Giardino}, P.~{Ferruit}, C.~{Alves de Oliveira}, S.~M.
  {Birkmann}, T.~{B{\"o}ker}, G.~{De Marchi}, X.~{Gnata}, J.~{K{\"o}hler},
  M.~{Sirianni}, and P.~{Jakobsen}.
\newblock {A model-based approach to the spatial and spectral calibration of
  NIRSpec onboard JWST}.
\newblock {\em Astronomy \& Astrophysics}, 592:A113, August 2016.

\bibitem{Norwood+others2016}
J.~{Norwood}, J.~{Moses}, L.~N. {Fletcher}, G.~{Orton}, P.~G.~J. {Irwin},
  S.~{Atreya}, K.~{Rages}, T.~{Cavali{\'e}}, A.~{S{\'a}nchez-Lavega},
  R.~{Hueso}, and N.~{Chanover}.
\newblock {Giant Planet Observations with the James Webb Space Telescope}.
\newblock {\em PASP}, 128(1):018005, January 2016.

\bibitem{Fehlmann+others2016}
A.~{Fehlmann}, C.~{Giebink}, J.~R. {Kuhn}, E.~J. {Messersmith}, D.~L. {Mickey},
  I.~F. {Scholl}, D.~{James}, K.~{Hnat}, G.~{Schickling}, and R.~{Schickling}.
\newblock {Cryogenic near infrared spectropolarimeter for the Daniel K. Inouye
  Solar Telescope}.
\newblock In {\em Society of Photo-Optical Instrumentation Engineers (SPIE)
  Conference Series}, volume 9908 of {\em Proc. SPIE}, page 99084D, August
  2016.

\bibitem{Johnson+others2014}
L.~C. {Johnson}, K.~{Cummings}, M.~{Drobilek}, S.~{Gregory}, S.~{Hegwer},
  E.~{Johansson}, J.~{Marino}, K.~{Richards}, T.~{Rimmele}, P.~{Sekulic}, and
  F.~{W{\"o}ger}.
\newblock {Solar adaptive optics with the DKIST: status report}.
\newblock In {\em Adaptive Optics Systems IV}, volume 9148 of {\em Proc. SPIE},
  2014.

\bibitem{VanDam+others2006}
M.~A. {van Dam}, A.~H. {Bouchez}, D.~{Le Mignant}, E.~M. {Johansson}, P.~L.
  {Wizinowich}, R.~D. {Campbell}, J.~C.~Y. {Chin}, S.~K. {Hartman}, R.~E.
  {Lafon}, P.~J. {Stomski}, Jr., and D.~M. {Summers}.
\newblock {The W. M. Keck Observatory Laser Guide Star Adaptive Optics System:
  Performance Characterization}.
\newblock {\em PASP}, 118:310--318, February 2006.

\end{thebibliography}

\end{document}

As an order of magnitude estimate of the effects of the denser
substrate on any large molecules and their spectra in the lab, the
molecular substrate can be considered as collisionless and
non-interacting for times of order
%$ \Lambda / \sqrt[]{kT/m({\rm H_2O})} \approx $ 
$10^{-2}$ seconds, using gas-kinetic cross sections of $\approx
10^{-16}$ cm$^2$.